\documentclass[preprint,10pt]{aastex}

\newcommand{\p}{$\pm$}

\newcommand{\beam}{beam$^{-1}$}

\newcommand{\mdot}{\.{M}}
\newcommand{\msun}{M$_{\sun}$}
\newcommand{\yr}{yr$^{-1}$}

\begin{document}

\title{Stellar Winds and Embedded Star Formation in the Galactic Center Quintuplet and Arches Clusters: Multifrequency Radio Observations}
\author{Cornelia C. Lang}
\affil{Department of Physics \& Astronomy, University of Iowa, Iowa City, IA 52245}
\email{cornelia-lang@uiowa.edu}
\and
\author{Kelsey E. Johnson\footnotemark\footnotetext{Hubble Fellow}}
\affil{Department of Astronomy, University of Virginia, Charlottesville, VA 22903}
\and
\author{W. M. Goss}
\affil{National Radio Astronomy Observatory, Socorro, NM 87801}
\and
\author{Luis F. Rodr\'{\i}guez}
\affil{Centro de
Radioastronom\'\i a y Astrof\'\i sica, UNAM, Campus Morelia, Apdo. Postal 3-72, Morelia, Michoac\'{a}n 58089, M\'{e}xico}

\begin{abstract}
A multi-frequency, multi-configuration study has been made of the compact radio sources in the Galactic Center 
Quintuplet and Arches stellar clusters using the Very Large Array. Ten radio sources have been
detected in the Quintuplet cluster. The majority of these radio sources have rising spectral indices
and are positionally coincident with young massive stars that are known to have powerful stellar winds. 
We conclude that the three most compact of these sources are produced by stellar wind emission; thus,
mass-loss rates can be derived and have an average value of 3$\times$10$^{-5}$ \msun~\yr. The remainder of 
the sources are likely to be a combination of stellar wind emission and free-free emission from surrounding ionized gas. 
In three cases, the radio sources have no stellar counterpart and the radio emission is thought 
to arise from compact or ultra-compact \ion H2 regions. If so, these sources would be the first 
detections of embedded massive stars to be discovered in the Galactic center clusters. The radio 
nebula associated with the Pistol star 
resembles the nebula surrounding the LBV star $\eta$ Carina and may be related to the stellar wind
of the Pistol star. Ten compact radio sources are detected in the Arches cluster and are interpreted to be stellar wind
sources, consistent with previous findings. Several of the sources show moderate 
variability (10-30\%) in their flux density, possibly related to a nonthermal component in the 
wind emission. A number of radio sources in both clusters have X-ray counterparts, which have been
interpreted to be the shocked, colliding winds of massive binary systems. 

\end{abstract}
\section{Introduction}

The Radio Arc near the Galactic Center (GC) ($\sim$30 pc in projection from the dynamical center of the Galaxy, Sgr A$^*$) is considered to be one of the most unusual collections of sources in the Galaxy~\citep{yzm84, yzm87a, myz89}. Both thermal and non-thermal sources comprise the Radio Arc. The most prominent sources are two sets of curved thermal filaments $-$ the Arched Filaments (G0.10+0.08) and the Sickle (G0.18$-$0.04)~\citep{yzm87s,myz89,lang97,lang01recomb}. These sources are all assumed to be at or very near the GC, which has a distance of 8.0 kpc; thus 1\arcsec~$\sim$0.04 pc (Reid 1993). One of the most exciting observational results of the last decade has been the near-infrared (near-IR) discovery of two extraordinary stellar clusters in the Radio Arc region, known as the Quintuplet and Arches clusters \citep{nagata95,cotera96,figerboth,figerquint}. These clusters have drastically changed our understanding of the stellar content in a region where the visual obscuration is 20$-$30 magnitudes. 

The Quintuplet cluster contains at least 10 Wolf-Rayet (WR) stars, and more than a dozen luminous ``OB supergiants'' \citep{figerboth,figerquint}. The unusual Pistol Star (luminosity of $\sim$10$^7$ L$_{\sun}$) is thought to be a member of the Quintuplet cluster \citep{figer95}. This star has been classified (along with one other Quintuplet member, \#362) as a Luminous Blue Variable (LBV) star~\citep{figerpistol,geballe00}. This type of evolved O-star (age $\sim$ 4 Myr) is thought to be undergoing short lived mass-loss during which up to several \msun~of material are ejected from its surface. The Arches cluster appears to be an even more extraordinary cluster with $\sim$35 WR stars as well as close to 100 luminous, evolved OB stars, making it one of the most densely-packed clusters in the Galaxy \citep{serabyn98,figerarches,blum01,stolte02}.  In total, nearly 50 stars have been classified as WR stellar types in the Radio Arc region \citep{cotera96,figerquint,figerarches,hom03}.

Such massive clusters are known to have a major impact on the interstellar medium. Figure 1 shows the locations of these two clusters relative to the main features in the Radio Arc. The Quintuplet and Arches clusters are responsible for the ionization of the Sickle and the Arched Filaments \ion H2 Regions \citep{cotera96,figerquint}. These luminous clusters with N$_{Lyc}$ $\sim$10$^{50.9}$ photons s$^{-1}$ (Quintuplet) and $\sim$4 $\times$ 10$^{51}$ photons s$^{-1}$ (Arches) can easily account for the average photon fluxes required from 8.5 GHz radio continuum observations (3-30 $\times$10$^{49}$ photons s$^{-1}$) \citep{lang97, lang01recomb}.

Radio emission from the individual stellar winds of such massive stars should be detectable at centimeter wavelengths based on the predictions of \citet{pf75} and \citet{wb75}. Indeed, high-resolution radio images made with the 
Very Large Array (VLA) of the National Radio Astronomy Observatory (NRAO)\footnotemark\footnotetext{The National Radio Astronomy Observatory is a facility of the National Science Foundation operated under cooperative agreement by Associated Universities, Inc.} have revealed a number of compact radio sources in these two clusters \citep{lang99,lang01winds}. Based on their spectral characteristics and positional associations with near-IR stellar sources, the radio sources are interpreted to be stellar wind emission in both clusters. The radio flux densities indicate that many of the sources have high mass-loss rates (\mdot~$>$ 2 $\times$10$^{-5}$ M$_{\sun}$ yr$^{-1}$). 

The initial results of the Quintuplet cluster radio sources presented by \citet{lang99} were based on low-resolution (2-10\arcsec) archival VLA data centered on the Sickle \ion H2 region and therefore were not ideal for detections of compact sources with sizes $<$1\arcsec. Here we present new, high resolution (0.3-3\arcsec) VLA images at 4.9, 8.5, 22.5 and 43.4 GHz of the Quintuplet and Arches stellar clusters that reveal new and variable radio sources and provide more accurate determination of the spectral indices and source sizes.  Although Lang et al. (2001b) presented the first sub-arcsecond imaging of the Arches cluster, here we provide a second epoch image at 8.5 GHz to search for new and variable radio sources in this cluster and an image of the cluster at 22.5 GHz to better characterize the spectral indices and sizes of the sources. The observations and data reduction are presented in $\S$2, the results in $\S$3, and the interpretation and discussion follow in $\S$4.

\section{Observations \& Imaging}

\subsection{Observations}
VLA multifrequency observations were made during 1999$-$2004 of the Quintuplet cluster, Pistol Star, and the Arches cluster using the A, BnA, B, D and DnC array configurations. Table 1 summarizes the observational details. At 4.9 and 8.5 GHz, J1331+305 and J1751-253 were used as flux and phase calibrators. Standard procedures for calibration, editing and imaging were carried out using the Astronomy Image Processing Software (AIPS) of the NRAO. At the higher frequencies (22.5 and 43.4 GHz), the observations were carried out differently. In the BnA and B array at both 22.5 and 43.4 GHz, the atmospheric phase fluctuations can be very pronounced, so that a fast cycle time between source and phase calibrator (Sgr A$^*$) was used, with a total cycle time of 2 minutes (90 seconds on source, 30 seconds on calibrator). The pointing was checked every hour using 8.5 GHz observations of 1751-253. In the DnC and D arrays at 22.5 and 43.4 GHz, a total cycle time of 5 minutes was employed (260 seconds on source, 40 seconds on calibrator). Pointing was also checked every hour by using 8.5 GHz observations of J1751-253. The absolute flux density scale was determined from the observations of J1331+305, with an assumed flux density of 2.52 Jy at 22.5 GHz and 1.45 Jy at 43.4 GHz and is used to measure a flux density of 1.3 Jy at 22.5 GHz and 1.4 Jy at 43.4 GHz for Sgr A$^*$. As the primary beam at 43.4 GHz is only 1\arcmin, the 43.4 GHz observations were centered on the extended sources QR1-3 (see Figure 2) rather than the center of the Quintuplet cluster since one of the goals of this study was to determine the flux densities of these unusual sources. 

\subsection{Imaging}

All images were created in AIPS using the standard imaging program IMAGR. Table 2 summarizes the final image resolutions and rms noise levels. The images shown in Figures 2-6 have not been corrected for primary beam attenuation, but flux densities were measured in images that had been corrected for this attenuation. Because the primary beams at 4.9 and 8.5 GHz are very large (9\arcmin~and 5.4\arcmin, respectively), and because the Quintuplet and Arches clusters are located near two very extended \ion H2 regions (see Figure 1), it is necessary to filter out part of the extended emission in the images. It is possible to do this with a radio interferometer by removing out the shortest {\it (u,v)} baselines in the {\it (u,v)} data, which correspond to the largest sized structures in the resulting image. Table 2 lists the cutoffs made to the data in the {\it (u,v)} plane (in kilo-wavelengths) and the corresponding size scales (in arcseconds) above which the flux is subsequently filtered out (i.e., the largest angular size scale present). In addition, at 22.5 GHz, two images of the Quintuplet cluster were produced: (a) a ``high'' resolution ($<$0.5\arcsec) image using {\it only} the B and BnA array data to investigate the most compact sources, and (b) a ``low'' resolution ($\sim$1.5\arcsec) image using all the available data to investigate the more extended sources.

\section{Results}
\subsection{Quintuplet Cluster}

Figure 2 shows the 8.5 GHz image of the radio sources within approximately 25\arcsec~($\sim$1 pc) of the Quintuplet cluster center. The Pistol nebula is a prominent extended radio source in this image and the near-IR position of the Pistol is represented by a cross. The Pistol nebula is believed to be the ejecta from a previous mass-loss event of the Pistol star, but also partly ionized by the Quintuplet cluster stars \citep{figerpistol,lang97,pistoleject}. Nine compact radio sources are detected and we refer to them as QR1-9. QR1-5 were previously identified by \citet{lang99}. QR6-9 are newly identified in this image. The radio sources were identified based on their signal-to-noise ratio (S/N $>$5 in most cases). The source QR9 has S/N $\sim$3 and is located outside of Figure 2, but was identified based on its positional association with a near-IR source (see $\S$3.1.2 below for more details on near-IR counterparts).

\subsubsection{Flux Densities, Spectral Indices and Source Structure}

Flux densities at all four frequencies (4.9, 8.5, 22.5 and 43.4 GHz) for QR1-9 and the Pistol Star were measured in primary beam corrected images and are listed in Table 3. The AIPS fitting routine JMFIT was used to measure the flux densities and deconvolved sizes for QR4-9.  The background levels in the area where QR1-3 are located (to the North of the Quintuplet cluster center; see Figure 2) vary strongly because of the proximity to the Radio Arc nonthermal linear filaments. In addition, the sources QR1-3 have lower surface brightnesses ($\sim$0.2 mJy \beam~at 8.5 GHz) than the more compact sources QR5-8 (average value of $\sim$0.6 mJy \beam~at 8.5 GHz), which makes the flux densities more difficult to calculate. In order to estimate accurate flux densities for these sources, the flux densities of several representative background regions were measured using the viewer tool in the {\it aips++} software package (http://aips2.nrao.edu). This tool was also used to integrate over the region of each source at all four frequencies simultaneously. The source flux densities were then corrected for the background levels at each frequency to determine the final flux densities listed in Table 3. 
Several sources were not detected at the higher frequencies. At 22.5 GHz, the source QR7 is not detected, most likely because of confusion in the vicinity of the Pistol nebula. Secondly, the source QR9 lies at the edge of the primary beam at 22.5 GHz, and attenuation likely prevents a detection. At 43.4 GHz, detections of sources QR4-8 were not made, probably due to the low resolution of the beam at this frequency ($\sim$2\arcsec), the fact that these sources are fairly compact ($\leq$1\arcsec), and the larger rms level in this image ($\sim$100 $\mu$Jy \beam). QR9 is not detected at 43.4 GHz as it lies outside of the 1\arcmin~primary beam. 

Spectral indices, $\alpha$, are calculated for all sources and are compiled in Table 4. Upper limits for flux densities at 4.9 GHz can be estimated for several of the sources not detected at this frequency, with resulting lower limits for the 4.9 to 8.5 GHz and 4.9 to 22.5 GHz spectral indices. QR1-3 and the Pistol Star show spectral indices consistent with being flat ($\alpha$$\sim$$-$0.3 to $-$0.1 with average errors of \p0.4). QR4-9 show rising spectral indices in the range of $\alpha$ = +0.4 to +2.0, also with large errors (\p0.2 to 0.8). Several sources exhibit complex structure, which makes determination of the spectral index difficult. Figure 2 (8.5 GHz image) and Figure 3 (the ``high'' resolution 22.5 GHz image) show that QR5 is a compact source. However, in the ``low'' resolution image at 22.5 GHz and in the 43.4 GHz image, this source has a compact center, but appears to have associated extended emission. Therefore, spectral indices are calculated for QR5 by comparing the flux densities in images where the source structure is similar (i.e., comparing the 8.5 GHz with the 22.5 GHz ``high'' resolution image and the 22.5 GHz ``low'' resolution image with the 43.4 GHz image).  QR7 is compact at 4.9 GHz, but fairly extended at 8.5 GHz and thus the spectral index is difficult to measure, but appears to be positive. QR9 is not detected 4.9 GHz and is detected only with a signal to noise ratio of 3.5 in the 8.5 GHz observations. We therefore estimate a lower limit for its spectral index of $\alpha$ $>$+0.2, implying that this source is likely to have a positive spectral index. 
 
The majority of the Quintuplet radio sources are resolved and the deconvolved sizes and corresponding linear sizes are presented in Table 3. The deconvolved sizes were derived from the 8.5 GHz image (Figure 2), although the sizes of QR5-9 were also measured in the ``high'' resolution image at 22.5 GHz where detected. The major axis sizes range from 0.3\arcsec~(0.015 pc) to 4.0\arcsec~(0.17 pc). Three sources (QR6, QR8, and QR9) are unresolved at both 8.5 and 22.5 GHz. For comparison, Figure 3 shows that QR5 is resolved and QR6 is unresolved in the 22.5 GHz ``high'' resolution image. Sources QR1-3 are more extended than the other sources in the Quintuplet cluster; their sizes range from 1.5\arcsec~(0.06 pc) to 4\arcsec~(0.16 pc). Figure 4 shows a detailed image of the sources QR1-3 at 8.5, 22.5 and 43.4 GHz.  

\subsubsection{Near-IR Counterparts}
The crosses in Figure 2 represent the positions of near-IR stars in the Quintuplet cluster from \citet{figerquint} with spectral classifications (representing only a fraction of the total number of luminous stars in the cluster). The positions of 28 stars are shown in Figure 2 (crosses) and represents the central region of the cluster. There is a systematic shift of positions between the near-IR and radio images ($\Delta$RA=0.9\arcsec), which is consistent with previous comparisons between {\it HST}/NICMOS images of the GC clusters and radio data \citep{lang99,lang01winds,figerarches}. All radio sources in the Quintuplet cluster appear to have near-IR counterparts within the positional uncertainty of 1\arcsec, except for QR1-3 (discussed later in $\S$4.1.3). Table 4 gives the stellar counterparts from Figer et al. (1999b). The near-IR counterparts to QR4, QR5, and the Pistol star were known from the results of Lang et al. (1999), but the near-IR counterparts of QR6-9 are newly associated.  The exact spectral classifications of these stars are not well known, although \citet{figerquint} provide some rough classifications in their Table 3. All the near-IR counterparts to the radio sources in our study appear to be evolving massive stars ($\sim$ few Myr), with many classified as WR stellar types (Figer et al. 1999a,b; Cotera et al. 1996). In particular, the near-IR counterpart of QR7 is one of five stars in the Quintuplet whose nature is unclear because the near-IR spectra are essentially featureless. \citet{figerquint} suggest that these five stars should be categorized as dusty, WC late-type stars (DWCL), which are the coolest carbon-rich WR stars that tend to form circumstellar dust shells. However, Moneti et al. (2001) present mid-infrared spectroscopic observations of these stars, but find no evidence (i.e., circumstellar absorption lines) to support the DWCL hypothesis. The near-IR counterpart of QR9 is also noteworthy as it has been classified as the second LBV star in the Quintuplet (\#362) besides the Pistol star \citep{figerquint,geballe00}.

\subsection{Pistol Star}
The radio source associated with the Pistol star is a prominent extended source located $\sim$10\arcsec~(0.4 pc) to the South of the Pistol nebula. A separate observation pointed in the direction of the Pistol star was made in 1999 in order to study the radio emission associated with this extraordinary star. Figure 5 shows that the radio source has a bipolar morphology, and that the Pistol star lies at the center of this source. The extended emission has a size of approximately 4\arcsec~(0.16 pc) in the N-S direction and approximately 2\arcsec~(0.08 pc) in the E-W direction. The flux densities for the Pistol radio source are summarized in Table 3. The spectral index of the radio emission is essentially flat ($\alpha_{22.5/8.5}$=0.0\p0.4). 

\subsection{Arches Cluster}
Figure 6 shows the 8.5 GHz image of the Arches cluster. Nine radio sources with signal-to-noise greater than three are present in the images and are labelled as AR1-9. AR10 is detected with a signal to noise ratio of only two, but is coincident with a near-IR counterpart. Sources AR1-8 were previously identified by \citet{lang01winds}, and AR9 and 10 are newly detected.  Five of these sources are also detected at 22.5 GHz.  

\subsubsection{Flux Densities \& Spectral Indices}
The flux densities at 8.5 and 22.5 GHz were measured for AR1-10 using the AIPS task JMFIT and are listed in Table 5. The 8.5 GHz observation presented here is a second epoch observation of this cluster and several sources showed moderate variability compared with the observations of \citet{lang01winds}: AR1, AR3, AR4 and AR8 varied by 12\%, 29\%, 30\%, and 25\%, respectively. The 22.5 to 8.5 GHz spectral index is calculated and also listed in Table 5. All of the sources except for AR6 show rising spectral indices.  AR1 is the only source that is resolved at 8.5 GHz, and has a deconvolved size of 0.14\arcsec$\times$0.10\arcsec~(0.006$\times$0.004 pc). At 22.5 GHz, AR1 is partially resolved with a deconvolved size of 0.12\arcsec~(0.005 pc) along its major axis. Although the 4.9 to 8.5 GHz spectral index for AR1 is rising ($\alpha$=+0.35\p0.04; Lang et al. (2001b), the flux density of this source at 22.5 GHz is less than at 8.5 GHz. It is also possible that some of this extended emission may be missing from the image of AR1 at 22.5 GHz as the largest angular size detected in the BnA array configuration of the VLA at 22.5 GHz could be as small as 2\arcsec. For this reason, the 22.5 to 8.5 GHz spectral index is not calculated for AR1. 

\subsubsection{Near-IR Counterparts}

The crosses in Figure 6 represent the positions of near-IR stars in the Arches cluster from \citet{figerarches}. Eight of these stars had been previously detected by \citet{nagata95} and \citet{cotera96} and classified with emission-line features. The recent observations of \citet{figerarches} confirm that at least thirty of the stars in the Arches cluster have the spectral features characteristic of young, massive stars, including all ten of the near-IR counterparts to the AR1-10 radio sources. Similar to the Quintuplet, there is a systematic shift ($\Delta$RA=0.8\arcsec, $\Delta$DEC=0.5\arcsec) between the near-IR and radio sources in the Arches cluster. However, this shift is within 1\arcsec, which is typical given the astrometric uncertainty of {\it HST}. The position of AR6 is found to be coincident with that of an emission line star (F19). There is also a weak radio source (AR10; 2$\sigma$ detection) associated with the near-IR source to the W of AR6 and its near-IR counterpart. 

\section{Discussion}
\subsection{Quintuplet Cluster: Nature of the Radio Sources}
\subsubsection{QR6, QR8 and QR9: Stellar Wind Sources}

Sources QR6, QR8, and QR9 show the major properties of stellar wind emission: (1) they have rising spectral indices in the range of +0.5 to +1.0, (2) they are unresolved sources at the resolutions observed, which means their physical sizes are $<$ 0.2\arcsec~(0.009 pc) at 22.5 GHz and $<$0.5\arcsec~(0.02 pc) at 8.5 GHz, and (3) they are coincident positionally with young, massive stars. Therefore, these three radio sources are interpreted to be stellar winds from Quintuplet cluster stars. 

The flux densities of radio detections of stellar winds can be used to calculate mass loss rates ($\dot{M}$). Based on Panagia \& Felli (1975) and Wright \& Barlow (1975), the mass loss rate can be estimated for an isotropic, homogeneous wind with constant velocity, electron density and chemical composition. However, 20-25\% of early-type stars are known to have a nonthermal (and variable) component in their winds (Abbott et al. 1984; Bieging et al. 1989).  The typical spectral index of such sources is in the range of $\alpha$$\sim$$-$0.6 to +0.3 for frequencies between 5 and 43 GHz (Beiging et al. 1989; Contreras et al. 1996). Contamination of the flux density by the nonthermal component can occur at the lower frequencies. Therefore, higher frequency observations (i.e., 23 GHz compared to 5 GHz) may be a better tracer of the thermal component and give more reliable mass-loss rates (Contreras et al. 1996). The mass-loss rate can be estimated from our 22.5 GHz observations using:  

\begin{equation}
\frac{\dot{M}}{10^{-5} M_\odot yr^{-1}}=0.34 \left(\frac{S_{22.5}}{mJy}\right)^{3/4} \left(\frac{v_{\infty}}{10^3~km~s^{-1}}\right) \left(\frac{d}{kpc}\right)^{3/2}, 
\end{equation} 
where S$_{22.5}$ is the flux density of the source at 22.5 GHz, v$_{\infty}$~is the terminal velocity of the wind (assumed to be 1000 km s$^{-1}$; Cotera et al. 1996; Nagata et al. 1995), and d is the distance to the source (8 kpc). We have assumed an electron temperature of 10$^4$ K, Z=1, and a mean molecular weight, $\mu$=2, due to the enrichment in heavy elements of the late-type WN stars (Leitherer et al. 1997), a classification which Figer et al. (1999) and Figer et al. (2002) has assigned to most of the Quintuplet and Arches cluster stars. Leitherer et al. (1997) suggest that there is very little variation in the mass-loss rates for WR stars, with an average value of $\sim$4 $\times$ 10$^{-5}$ \msun~\yr. Cappa et al. (2004) recently carried out a survey of Galactic WR stars and came to similar conclusions concerning the measured radio mass-loss rates; the average values from their study were 2-4 $\times$ 10$^{-5}$ \msun~\yr. Values for the mass-loss rates of QR6, QR8 and QR9 range from 3.2$\times$10$^{-5}$ \msun~\yr~to 4.3$\times$10$^{-5}$ \msun~\yr~ (see Table 4). These values agree well with those found for WR stars described above as well as for other young, massive stellar types. 

\subsubsection{Sources QR4, QR5 and QR7: Stellar Wind and \ion H2 Emission?}

The nature of radio emission from sources QR4, QR5 and QR7 is less clear than that from QR6, QR8 and QR9. Although QR4, QR5 and QR7 show rising spectra and are coincident with near-IR sources, their sizes are more extended ($>$1\arcsec~at 8.5 GHz) than expected for stellar wind sources. According to Rodriguez et al. (1980), the angular size of a stellar wind can be related directly to its flux density, independent of mass-loss rate and terminal wind velocity. For the 8.5 GHz flux densities of QR4, QR5 and QR7, the expected angular size of the corresponding stellar wind should be $<$0.05\arcsec. The actual sizes of the sources (0.3\arcsec~to 4\arcsec) are clearly larger than the predicted value. However, as mentioned previously (in $\S$2.2), the Quintuplet radio sources are located amidst a very complicated set of extended features, which includes the Sickle and Pistol \ion H2 regions as well as the linear, nonthermal filaments of the Radio Arc. Although filtering in the {\it (u,v)} plane can help by removing the extended structures, the Quintuplet sources may still be contaminated by residual background emission, which can make determination of flux densities and sizes difficult. Although the errors in size determination are substantial (0.3$-$0.5\arcsec; 20-50\%), the sizes are still much more extended than the predicted values for stellar winds (e.g., $<$0.05\arcsec). Mass-loss rates are calculated for these sources as described above, and the values for QR4, QR5 and QR7 (average \mdot~of 1.2 $\times$ 10$^{-4}$ \msun~\yr) are significantly larger than for the stellar wind sources QR6, QR8 and QR9 (average \mdot~of 3.7$\times$10$^{-5}$ \msun~\yr).  While values for \mdot~of $\sim$ 10$^{-4}$ \msun~\yr~are not unexpected for some mass-losing stars, they typically correspond to epochs when the star may be experiencing an enhanced rate of mass-loss from its surface (i.e., the LBV phase; White et al. 1994). 

Therefore, the simple interpretation of stellar wind emission may not apply for QR4, QR5 and QR7.  However, the rising spectra and extended sizes of these sources are consistent with partially-thick free-free emission associated with an inhomogeneous, but spatially unresolved, \ion H2 region. Therefore, QR4, QR5 and QR7 could be inhomogeneous \ion H2 regions or even combinations of stellar wind emission and optically-thin, free-free emission from associated ionized material (perhaps leftover from the formation of the stars). Based on Mezger \& Henderson (1967), we can estimate the physical properties of these regions of ionized gas, assuming they have become optically-thin at the highest frequency at which they are detected, using the deconvolved sizes of these sources, and using an electron temperature of 10$^4$ K. Table 4 lists the electron density, n$_e$, mass of ionized hydrogen, M$_{HII}$, emission measure, EM, and the production rate of ionizing photons, N$_{Lyc}$ for these sources. The values of these \ion H2 properties for QR4, QR5 and QR7 (in addition to their linear sizes of 0.03-0.05 pc) are consistent with those for compact and ultra-compact \ion H2 regions (Wood \& Churchwell 1990). 

\subsubsection{QR1, QR2 and QR3: Embedded Star Formation in the Quintuplet Cluster}

The properties of QR1-3 differ from the other radio sources in the Quintuplet. As described in $\S$3.1.1, QR1-3 are more spatially extended than QR6, QR8 and QR9 by at least a factor of 2. More significantly, QR1-3 are the only radio sources in the two clusters without closely associated near-IR counterparts. These sources also define the northern edge of the Quintuplet cluster (5-10\arcsec~north of QR4-7 and the main concentration of stellar members). Table 4 shows that the spectral indices are close to flat with large errors (i.e. $\alpha$=$-$0.3\p0.2 for $\alpha_{22.5/8.5}$ for QR1). Such spectra can be attributed to optically thin, free-free emission associated with \ion H2 regions. Since there are no stellar counterparts to QR1-3, a likely explanation for these sources is that they may represent embedded massive stellar sources (Churchwell 1990). The linear sizes of QR1-3 of $\sim$0.1 pc are consistent with the upper end of the size distribution of ultra-compact \ion H2 regions in the sample of Kurtz, Churchwell, \& Wood (1994).  As only $\sim 50$\% of ultra-compact \ion H2 regions are detectable at near-IR wavelengths (Hanson, Luhman, \& Rieke 2002), the lack of near-IR emission associated with QR1-3 is also consistent with their being ultra-compact \ion H2 regions. 

As described above, the \ion H2 properties for QR1-3 are derived from the radio continuum and listed in Table 4. 
The production rate of ionizing photons, N$_{Lyc}$, can provide an estimate of  
the embedded stellar type that is responsible for the associated radio
emission (assuming a single ionizing star). Values for N$_{Lyc}$ range from $\approx 1-2 \times 10^{46}$~s$^{-1}$.  
These ionizing fluxes are consistent with embedded B-type massive stars  
($\sim$ B1V-B2V; \citep{smith02}).  However, a number of Galactic 
ultra-compact \ion H2 regions are observed to be associated with diffuse extended emission 
\citep{kim01, kurtz99}.  These studies
conclude that typically $\gtrsim 80$\% of the ionizing flux from the
embedded stars in UC \ion H2s escapes to the outer envelope.
Thus, the inferred stellar content from the $N_{Lyc}$ values
determined for sources QR1-3 may be an underestimate, in which case
they could have spectral types earlier than B. 
Therefore, we tentatively identify QR1-3 as ultracompact or compact \ion H2 regions.

The proximity of the Quintuplet cluster radio sources (and in particular, sources QR1-3) 
to the edge of a molecular cloud identified by Serabyn \& G\"{u}sten (1991) 
may indicate that these sources are physically associated with this molecular cloud.  
In the CS (J=3-2) image of Serabyn \& G\"{u}sten, a
finger of emission extends toward the projected location of QR1-3,
indicating the presence of moderately dense gas.  The apparent
physical association between QR1-3 and the edge of the molecular cloud
suggests that these young stellar objects have recently
emerged from the molecular cloud, probably as a result of erosion due
to the more evolved massive stars in the Quintuplet cluster. Indeed,
the Quintuplet cluster is thought to be the source of
ionization for the larger Sickle \ion H2 region (Lang, Goss \& Wood 1997; Figer et al. 1999a), 
which traces the inner edge of the molecular cloud.

Alternatively, the radio sources which are not interpreted to be stellar winds 
(QR1-3, QR4, QR5, and QR7) may represent clumps of molecular material leftover from the
formation of the stars in the cluster that are externally ionized
by the Quintuplet cluster members. Assuming these clumps are located at approximately 
the projected distances from the center of the cluster, the
Quintuplet could easily account for their ionization as it is responsible for
the ionization of the much larger Sickle \ion H2 region (located 1\arcmin~or 2.5 pc to the NE)
and the north side of the Pistol nebula (Lang et al. 1997; Figer et al. 1998). 
If any of the Quintuplet radio sources are confirmed to be compact or UC \ion H2s, this
detection will lead to the study of the earliest stages of stellar evolution
in one of the most extreme environments in the Milky Way.  In
particular, these objects are located in a region of strong magnetic fields, high
turbulent velocities, and extremely high metallicities (Morris \& Serabyn 1996). 

\subsubsection{Extended Emission Surrounding the Pistol Star}

Figure 5 shows that the radio emission associated with the Pistol Star has a bipolar morphology centered around the position of the star and has a flat spectral index ($\alpha_{22.5/8.5}$ = +0.1\p0.4). One explanation for this source is that it represents the current mass-loss of the star. (The outer ``Pistol-shaped'' nebula is thought to contain the mass associated with an episode of previous mass-loss ($\sim$~11 \msun, $\sim$5000 years ago; Figer et al. 1998)). In this case, the integrated flux density would imply a mass-loss rate of 2.7$\times$10$^{-4}$~\msun~\yr. This value is much higher than the average value for WR stellar types (see $\S$4.1.1), but similarly high mass-loss rates have been detected in other LBV stars during epochs of extreme mass-loss (e.g., $\eta$ Car; White et al. 1994) and in several WR colliding-wind binary systems (Contreras et al. 1996). 

However, as for sources QR4, QR5, and QR7, the extended size of the Pistol Star radio emission is much larger 
than the expected angular size 
for a stellar wind source ($\sim$0.1\arcsec) using the methods of Rodr\'{\i}guez et al. (1980).    
Another problem with the interpretation of the Pistol as a stellar wind source is that its
double-lobed radio emission is not well modelled by a spherically-symmetric stellar wind surrouding
the star. It is possible to calculate a mass-loss rate for the source using the total mass and assuming
an expansion velocity for the radio emission.  Assuming that the radio emission 
is optically-thin, free-free emission, with T$_e$$\sim$10$^4$ K, 
and taking the deconvolved angular sizes of the lobes to be 1$''$,
we can estimate the total mass in the lobes to be $\sim$ 0.02 $M_\odot$.
The separation between the Pistol star and each of the
lobes is $\sim 1.2 \times 10^{17}$ cm. Assuming that the lobes
are expanding at a velocity of 1000 km s$^{-1}$, we estimate a mass-loss rate of $5 \times 10^{-4}$ $M_\odot$ yr$^{-1}$. This value is roughly consistent with the value obtained by using equation (1). 
In addition, it is possible to calculate the ionizing flux required to maintain the bipolar radio
source. At a distance of 8 kpc and with the integrated flux density measured, N$_{Lyc}$ has a value of
 $\sim 2 \times 10^{46}$ s$^{-1}$, which can be provided by an early B-type star. 
However, the two lobes that constitute the radio source
are intersecting only a fraction of the 4$\pi$ radians around the
star. Furthermore, the ionized nebula may be density-bounded, with most
of the ionizing flux of the Pistol star escaping to the interstellar medium, so the star is likely
to be of an earlier spectral type.

The physical properties of the radio emission associated with the Pistol star are similar to the nebula surrounding the well-known LBV binary system $\eta$ Carina (Davidson \& Humphreys 1997).  The bipolar-shaped radio emission around $\eta$ Carina has a flat spectral index and a linear size of 0.12 pc (10\arcsec~at 2.5 kpc; White et al. 1994), similar to the Pistol star nebula. In fact, a variety of types of nebular features are detected around other LBV stars and are thought to be the result of stellar wind emission interacting with the surrounding interstellar medium (Nota et al. 1995). In the cases of Galactic sources HG Car, AG Car, WRA-751 and He3-519, both extended nebular shells and compact stellar wind emission are detected in the radio continuum (Duncan \& White 2002). These authors also calculate substantial mass-loss rates from the central LBV stars ($>$5-10 $\times$ 10$^{-5}$ \msun~\yr). The bipolar morphology of such radio nebulae suggests that the mass-loss is not isotropic (e.g., $\eta$ Car). This type of non-spherical wind emission has been explained by several different mechanisms: (1) {\it internal} processes, such as the stellar rotation of such massive stars and the resulting mass-loss geometry (Owocki \& Gayley 1997; Maeder \& Maynet 2000; Smith et al. 2004), and (2) {\it external} processes, such as the influence of a binary companion on the mass-loss geometry (Iben \& Livio 1993) or expansion of the nebula into a region of high density-contrast interstellar medium (Nota et al. 1995). 

\subsection{Arches Cluster}

A comparison of the 8.5 GHz observation of the Arches cluster presented here and the observation of Lang et al. (2001b) 
shows that the majority of the
sources do not exhibit any strong variability in their flux density. However, moderate variability 
(at the level of 12-30\%) is detected in 
several of the Arches radio sources previously detected (AR1, AR3, AR4, AR8) and possibly explains
the new detection of AR9. The flux densities of two of the four sources decreased, while the
flux density of the other two sources increased. Therefore, we can rule out a systematic 
calibration problem. This moderate variability may indicate (i) the presence of a 
time-variable nonthermal component (due to shocks in the wind, or colliding
 winds for a binary system) and/or (ii) variability in the mass-loss rate or wind velocity
(Contreras et al. 1996; Leitherer et al. 1997). 

Nine of the ten Arches sources are unresolved in both the 8.5 and 22.5 GHz observations. The major axis of
AR1 is resolved at both 8.5 and 22.5 GHz and its deconvolved size is $\sim$0.1\arcsec. This size is consistent with the expected
angular size of 0.06\arcsec~for an ionized stellar wind source of 1.9 mJy flux density at 8.5 GHz (Rodr\'{\i}guez et al. 1980). Therefore, all radio sources in the Arches cluster exhibit the properties of stellar wind sources (see $\S$4.1.1).  
Mass-loss rates are also calculated for the Arches sources and are listed in Table 5. AR1 has the largest value of 
\mdot~=2.2 $\times$ 10$^{-4}$~\msun~\yr. Such a high value of \mdot~is usually attributed to a particular mass-loss
epoch during the evolution of a massive star or a colliding wind binary system (White et al. 1994; Contreras et al. 1996). 
The remainder of the Arches sources have an average value of $\sim$3 $\times$ 10$^{-5}$~\msun~\yr, 
typical for the types of stars present in the Arches and Quintuplet clusters 
($\S$4.1.1; Leitherer et al. 1997; Cappa et al. 2004). 

\subsection{X-Ray Counterparts in the Arches and Quintuplet Clusters}

As reported recently by Law \& Yusef-Zadeh (2004) and previously by Yusef-Zadeh et al. (2002), a number of 
the radio sources in the Arches and Quintuplet clusters have X-ray counterparts detected with the {\it Chandra X-Ray Observatory}. The X-ray emission associated with the Arches cluster itself is one of the most luminous X-ray sources in the Galactic Center region with an X-ray luminosity in the 2-10 keV range of $\sim$10$^{35}$ ergs s$^{-1}$ (Wang, Gotthelf \& Lang 2002). Three X-ray point like sources are detected in the Arches cluster: A1S, A1N and A2 (Law \& Yusef-Zadeh 2004). The X-ray sources A1S and A1N can be associated with the radio sources AR4 and AR1 as well as near-IR sources from Figer et al. (2002). The X-ray emission in these sources is thought to arise from shocks formed in the winds of massive stars (O- and WR-types) as are found in the Arches and Quintuplet clusters (Law \& Yusef-Zadeh 2004). 

The characteristics of several Arches radio sources that have X-ray counterparts suggest that these sources are likely to have shocks in their winds and/or be colliding wind binaries. Radio emission from the colliding winds of binary massive stars is well known to have a nonthermal component that flattens the spectral index (Dougherty \& Williams 2000) as well as showing time-variable emission (Abbott et al. 1984). In the case of the radio source AR1, the spectral index ($\alpha$=+0.3\p0.04; Lang et al. 2001b) is flatter than the canonical $\alpha$=+0.6 value for stellar wind emission and the flux density of the source varies by 12\% between epochs. AR4 has a steeper spectral index (although the errors are larger), $\alpha$=+0.6\p0.2, but its flux density varied by 30\% in the two observed epochs. 

In the Quintuplet cluster, Law \& Yusef-Zadeh (2004) report that two of the radio sources (QR6 and QR7) have probable X-ray counterparts. It is not possible to determine whether the spectral index of QR6 is flattened ($\alpha$=+0.4\p0.4) or consistent with the value of $\alpha$=+0.6. However, this source was not detected in the results of Lang et al. (1999), either because of its time-variable nature or because the resolution was too coarse. The possible detection (2.6$\sigma$ as reported by Law \& Yusef-Zadeh 2004) of X-ray emission associated with Quintuplet source QR7 is interesting as not much is known about the nature of the near-IR counterpart ($\S$3.1.2; Figer et al. 1999). 

\subsection{Impact of Massive Star Winds on the GC Interstellar Medium}

The Arches and Quintuplet clusters are two of the most massive and luminous clusters in our Galaxy (Figer et al. 1999a). In particular, the stellar density of the Arches cluster ($\rho$ $\sim$ 3 $\times$10$^5$ \msun~pc$^{-3}$) is comparable only to that found in the centers of clusters such as HD in NGC3603 (Moffat et al. 1994) and R136 in the Large Magellanic Cloud, associated with the 30 Doradus nebula (Massey \& Hunter 1998). Therefore, these clusters should have a profound effect on the surrounding ISM. Indeed, these clusters are thought to be responsible for the ionization of the adjacent \ion H2 regions, the Sickle and the Arched Filaments (Lang et al. 1997; Lang et al. 2001a). 

The presence of X-ray emission associated with individual stellar wind sources (see above) also indicates that the winds of these massive stars are powerful. In addition to the point-like X-ray sources, both clusters also show diffuse X-ray emission (Yusef-Zadeh et al. 2002; Law \& Yusef-Zadeh 2004). In the Arches cluster, the diffuse X-ray emission is apparent for $\sim$1 pc surrounding the cluster, and in the Quintuplet, the diffuse X-ray emission has a slightly larger extent ($\sim$1.2 pc) but is not detected with high signal to noise (Law \& Yusef-Zadeh 2004). Several authors have suggested that the diffuse emission is related to a ``cluster wind'' triggered by the collective effect of a large number of massive star winds in the cluster core (Cant\'{o} et al. 2000; Raga et al. 2001; Yusef-Zadeh et al. 2002). Such processes can lead to shock acceleration of the wind particles and heating of the gas to high temperatures (10$^7$ to 10$^8$ K). Nonthermal radio emission has been detected within a 9\arcsec~radius (0.3 pc) of the cluster core and is believed to provide additional evidence that particle acceleration occurs here (Yusef-Zadeh et al. 2003).  

The physical parameters of the ``cluster wind'' and the resulting region of X-ray emitting gas depend primarily on the individual mass-loss rates of the stellar wind sources (Stevens \& Hartwell 2003; Cant\'{o} et al. 2000; Raga et al. 2001). The total stellar wind mass injection rate is dependent on the sum of mass-loss rates of individual stars multiplied by the square of the terminal wind velocity. Using an estimate for the velocity, predicted by the models in Stevens \& Hartwell (2003), or by using the average value of $\sim$1000 km s$^{-1}$, measured by Cotera et al. (1996), it is possible to derive the total stellar wind kinetic energy injection rate.  The observations presented here provide the most reliable mass-loss rate estimates for the stellar wind sources in the Arches and Quintuplet clusters. Based on the mass-loss rates presented here, the total stellar wind mass injection rate for the Arches cluster is $\sim$4 $\times$ 10$^{-4}$ \msun~\yr, and for the Quintuplet (including the Pistol star as well as sources QR4, QR5, and QR7) is $\sim$7 $\times$ 10$^{-4}$ \msun~\yr.  These values for total \mdot~are in close agreement with those used by Stevens \& Hartwell (2003) and Raga et al. (2001) in their calculation of the total stellar wind kinetic energy injection rate.

Two types of numerical calculations have also been made recently to characterize the X-ray ``cluster wind'' that arises from the Arches (and presumably, the Quintuplet) clusters. Cant\'{o} et al. (2000) and Raga et al. (2001) present a solution for an adiabatic flow and predict an envelope of X-ray emission of T$\sim$5 $\times$10$^7$ K, extending well beyond (10's of pc) the cluster core (0.2 pc). In contrast, Silich et al. (2004) employ radiative cooling in their work, and predict, for the Arches cluster, a quasi-adiabatic solution that suggests a much smaller boundary around the cluster of X-ray emission. Depending on the values chosen for \mdot~and the terminal velocity, Silich et al. (2004) estimate 3-7 pc for the extent of the X-ray emitting region. These authors argue that this smaller size for the X-ray emitting region is in part due to the slowing of wind emission by interactions of the wind with cooler surrounding ISM material. The Arches cluster, in particular, is located in the midst of a dense molecular and ionized environment (Lang et al. 2002) and there is a large reservoir of cooler material which may decelerate the wind. Morphologically, the Arched Filaments \ion H2 regions and molecular cloud extend for as much as 20 pc in projection around the cluster. X-ray observations near the Arched Filaments show that the diffuse emission appears to be confined to the area near the easternmost of the Arched Filaments ($<$ 10 pc), as predicted by Silich et al. (2004).  However, the short-integration nature of these observations ($<$ 50 ks) with {\it Chandra} and the subsequent low sensitivity to diffuse objects have so far prevented detailed spectral fitting and spatial imaging of the diffuse X-ray emission associated with these clusters (Yusef-Zadeh et al. 2002; Wang, Gotthelf \& Lang 2002). 

\subsection{Clusters of Radio Sources Associated with Regions of Recent Massive Star Formation}

Several examples of radio clusters associated with regions of recent star formation have been reported in the 
literature over the last ten years
(e.g., Garay et al. 1987; Churchwell et al. 1987; Becker \& White 1988;
Stine \& O'Neal 1998; Rodr\'\i guez et al. 1999; G\'omez et al. 2000; Lang et al. 2001).
The compact radio sources in the stellar clusters have a variety of origins: 
ionized stellar winds, ultracompact \ion H2 regions,
externally ionized proplyds, and magnetically active stars with 
gyrosynchrotron emission are the dominant types of sources. 
In Table 6 we summarize the parameters of the radio sources in these clusters.
Although more of these clusters should be studied to have a reliable statistical
base, some interesting trends are evident. The diameters of the radio clusters are in the
0.2 to 1.0 pc range (with the exception of the Quintuplet cluster with the Pistol star
included, that reaches up to 2 pc). 
The radio luminosity of the most luminous member is correlated with
the bolometric luminosity of the cluster.
Clearly, the Arches and Quintuplet cluster have their most luminous radio
member with luminosities one to three orders of magnitude larger than
typical young clusters in the Galactic disk. 
These clusters of radio sources could be a valuable observational
tool to study the stellar population of heavily obscured regions of star formation when combined with 
high resolution spectroscopic observations.

\section{Conclusions}

We have made multi-frequency radio observations of both the Quintuplet and Arches clusters using the VLA. Our conclusions can be summarized as follows: 

(1)  Ten radio sources have been detected in the Quintuplet cluster. The majority of these sources have rising spectral indices and are positionally coincident with near-IR sources which have been classified as young, massive stars. Sources QR6, QR8 and QR9 are unresolved and thought to be very compact in size ($<$0.01 pc); therefore, these three sources are interpreted to be stellar winds. Sources QR4, QR5 and QR7 are more extended (0.3$-$1.5\arcsec; 0.015 to 0.06 pc) and likely represent a combination of stellar wind and free-free emission associated with \ion H2 regions. Sources QR1-3 do not have near-IR counterparts and these are interepreted to be the first detections of embedded massive stars in the Quintuplet cluster.   

(2) An observation of the Pistol star shows that its structure is extended and distinctly bipolar, centered on the Pistol star near-IR source. The radio emission from this nebula has a flat spectral index. Assuming that this emission is optically thin, free-free emission, we can roughly calculate a mass-loss rate of 5 $\times$ 10$^{-4}$ \msun~\yr. The Pistol star radio source resembles the nebula that surrounds the luminous blue variable star $\eta$ Carina.

(3) Ten compact radio sources have also been detected in the Arches cluster. Nine of the sources have spectral indices consistent with a rising spectrum and only one of the sources may be resolved. As found in Lang et al. (2001b), these sources are interpreted to be stellar winds from the young massive stars in the cluster. The average mass-loss rate for these stellar winds is 3 $\times$ 10$^{-5}$ \msun~\yr.  

(4) Both point-like and diffuse X-ray emission has been identified with the Arches and Quintuplet clusters and thought to be related to the presence of colliding wind binaries in these clusters. The mass-loss estimates from our radio observations here provide the best estimates of one of the parameters used to model the X-ray ``cluster winds'' which are observed to be associated with a number of massive stellar clusters.  

C.C.L. acknowledges that this material is based on work supported by the National Science Foundation (NSF) 
under Grant No. 0307052. K.E.J gratefully acknowledges support for this work provided by the NSF through 
an Astronomy \& Astrophysics Postdoctoral Fellowship,
and NASA through Hubble Fellowship grant \#HF-01173.02-A awarded by the
Space Telescope Science Institute, which is operated by the Association
of Universities for Research in Astronomy, Inc., for NASA, under
contract NAS 5-26555.

\begin{deluxetable}{lcccccc}
\tabletypesize{\scriptsize}
\tablecaption{Observational Properties\label{observations}}
\tablehead{\colhead{Source} &
\multicolumn{2}{c}{Pointing Center}&
\colhead{Frequency}&
\colhead{Array}&
\colhead{Observation}&
\colhead{Integration}\\
\cline{2-3}
\colhead{Name}&
\colhead{R.A. (J2000)} &
\colhead{Decl. (J2000)} &
\colhead{(GHz)} &
\colhead{Config.} &
\colhead{Date}&
\colhead{Time (hrs)}}
\tablecolumns{7}
\startdata
Quintuplet&17 46 15.00 &$-$28 49 35.0&4.9 &BnA&June 2002& 4 \\
Quintuplet&17 46 15.00 &$-$28 49 35.0&8.5 &BnA&June 2002& 4 \\ 
Quintuplet&17 46 15.00 &$-$28 49 35.0&22.5 &BnA&Sept 2003&2 \\
Quintuplet&17 46 15.00 &$-$28 49 35.0&22.5 &B&Jan. 2004&3 \\
Quintuplet&17 46 15.00&$-$28 49 35.0&22.5 &DnC,D&June 2004&3.5 \\
QR1-3 Sources&17 46 14.60&$-$28 49 20.0&43.4 &DnC&June 2004&2.5 \\
QR1-3 Sources&17 46 14.60&$-$28 49 20.0&43.4 & D& July 2004&3 \\
Pistol & 17 46 15.24 & $-$28 50 04.0 & 8.5 & B, BnA & Nov. 1999 & 4 \\
\hline
Arches&17 45 50.41 & $-$28 49 21.8 &8.5 &A&Jan. 2002&4\\
Arches&17 45 50.41 & $-$28 49 21.8 &22.5 &BnA&Sept. 2003& 2.5\\
\enddata
\end{deluxetable}

\begin{deluxetable}{lclccc}
\tabletypesize{\scriptsize}
\tablecaption{Image Properties\label{images}}
\tablehead{\colhead{Field} &
\colhead{Frequency}&
\colhead{Resolution}&
\colhead{RMS Noise}&
\colhead{{\it (u,v)} cutoff}&
\colhead{Largest Angular}\\
\colhead{Name}&
\colhead{(GHz)}&
\colhead{(\arcsec$\times$\arcsec)}&
\colhead{($\mu$Jy \beam)}&
\colhead{(kilo-$\lambda$)}&
\colhead{Scale (\arcsec)}}
\tablecolumns{6}
\startdata
Quintuplet& 4.9 &1.15 $\times$ 0.85, PA=62.8\arcdeg&75&10&20.6\\
Quintuplet& 8.5 &0.64 $\times$ 0.50, PA=62.0\arcdeg&55&10&20.6\\
Quintuplet&22.5 &0.39 $\times$ 0.22, PA=52.7\arcdeg\tablenotemark{\dagger}&65&15&13.8\\
Quintuplet& 22.5 &1.45 $\times$ 0.87, PA=06.8\arcdeg\tablenotemark{\dagger\dagger}&75&$-$&$-$\\
Quintuplet (QR1-3)& 43.4 & 2.54 $\times$ 1.35, PA=03.3\arcdeg&100&$-$&$-$\\
Pistol &  8.5 & 1.17 $\times$ 0.59, PA=15.2\arcdeg&75&25&8.2\\
\hline
Arches&8.5 & 0.42 $\times$ 0.20, PA=01.6\arcdeg&25&20&10.3\\
Arches&22.5 &0.33 $\times$ 0.20, PA=47.8\arcdeg&85&10&20.6\\
\enddata
\tablenotetext{\dagger}{``high'' resolution image, made with B and BnA array data.}
\tablenotetext{\dagger\dagger}{``low'' resolution image, made with BnA, B, DnC and D array data.}
\end{deluxetable}

\begin{deluxetable}{lcccccccc}
\tabletypesize{\scriptsize}
\tablecaption{Observed Properties of the Radio Sources in the Quintuplet Cluster}
\tablehead{
\colhead{Source} &
\multicolumn{2}{c}{Source Position\tablenotemark{a}}&
\multicolumn{4}{c}{Flux Density (mJy)\tablenotemark{b}} &
\colhead{Deconvolved}&
\colhead{Linear}\\
\cline{2-3}
\cline{4-7}
\colhead{Name}&
\colhead{R.A. (J2000)} &
\colhead{Decl. (J2000)} &
\colhead{4.9 GHz} &
\colhead{8.5 GHz} &
\colhead{22.5 GHz}&
\colhead{43.4 GHz}&
\colhead{Size (\arcsec)\tablenotemark{c}}&
\colhead{Size (pc)\tablenotemark{d}}}
\tablecolumns{9}
\startdata
QR1	&17 46 15.16	&$-$28 49 22.5	& 2.2\p0.4& 1.9\p0.2&1.4\p0.3&1.0\p0.5&2.5$\times$1.5& 0.10$\times$0.06\\
QR2	&17 46 14.95	&$-$28 49 20.2	& 3.6\p0.7& 3.4\p0.3& 2.4\p0.4	&1.4\p0.6&4.0$\times$2.5& 0.17$\times$0.10\\
QR3	&17 46 14.42	&$-$28 49 19.4	& 1.7\p0.4& 1.5\p0.2& 1.1\p0.3 & 0.3\p0.2&2.5$\times$2.5& 0.10$\times$0.10\\
QR4	&17 46 15.06	&$-$28 49 29.4	&0.60\p0.30 &1.30\p0.30  & ...	& ... &1.4$\times$0.9&0.04$\times$0.03\\
QR5	&17 46 15.11	&$-$28 49 37.2	&0.35\p0.10 &0.54\p0.10  & 1.50\p0.20	& ... & 0.3$\times$0.3\tablenotemark{e} &0.01$\times$0.01\\
QR5ext\tablenotemark{f}...	& ...	& ...	&... & ...  & 2.60\p1.00	& 2.80\p1.6& 1.6$\times$1.2& 0.06$\times$0.05\\
QR6	&17 46 15.14	&$-$28 49 32.7	&0.25\p0.10 &0.33\p0.10  & 0.48\p0.08     &...&ps &...\\
QR7	&17 46 14.71	&$-$28 49 40.8	&$<$0.24 &0.80\p0.20  & ...  & ... &1.3$\times$0.7&0.05$\times$0.03 \\
QR8	&17 46 14.04	&$-$28 49 16.7	&$<$0.24    &0.28\p0.10  & 0.57\p0.08	&...&ps & ...\\
QR9	&17 46 17.98	&$-$28 49 03.4	&$<$0.24    &0.27\p0.06  & ...	& ... &ps & ...\\
Pistol  &17 46 15.32    &$-$28 50 04.2  &3.2\p1.0  & 3.4\p1.0&3.5\p1.0 &3.4\p1.0 &4.0$\times$2.5 &0.17$\times$0.10\\
\enddata
\tablenotetext{a}{Positional errors are \p0.01 in RA and \p0.1\arcsec~in DEC.}
\tablenotetext{b}{Flux densities were measured from images corrected for primary beam attenuation.}
\tablenotetext{c}{Deconvolved sizes measured at 8.5 or 22.5 GHz; ps=point source}
\tablenotetext{d}{Linear sizes calculated assuming a distance to the GC of 8.0 kpc (Reid et al. 1993).}
\tablenotetext{e}{As measured in the ``high'' resolution 22.5 GHz image (Figure 3).}
\tablenotetext{f}{Extended emission present around QR5 in the ``low'' resolution 22.5 GHz image.}
\end{deluxetable}

\begin{deluxetable}{lccccccccccc}
\tabletypesize{\tiny}
\tablecaption{Derived Properties for the Quintuplet Cluster Sources}
\tablehead{
\colhead{Source} & 
\multicolumn{4}{c}{Spectral Index\tablenotemark{a}}&
\colhead{\mdot} &
\colhead{n$_e$}&
\colhead{M$_{HII}$}&
\colhead{EM}&
\colhead{N$_{Lyc}$}&
\colhead{Near-IR} \\ 
\cline{2-5}
\colhead{Name}&
\colhead{$\alpha_{8.5/4.9}$}&
\colhead{$\alpha_{22.5/8.5}$}&
\colhead{$\alpha_{22.5/4.9}$}&
\colhead{$\alpha_{43.4/22.5}$}&
\colhead{(M$_\sun$ yr$^{-1}$)\tablenotemark{b}}&
\colhead{(cm$^{-3}$)}&
\colhead{(M$_{\sun})$}&
\colhead{(pc cm$^{-6}$)} &
\colhead{(photons s$^{-1}$)}&
\colhead{Counterpart\tablenotemark{c}}}
\tablecolumns{11}
\startdata
QR1	& $-$0.3\p0.4&$-$0.3\p0.2 &$-$0.3\p0.3&$-$0.5\p0.8 & $-$ & 1.5$\times$10$^3$& 0.023&2.4$\times$10$^5$&1.2$\times$10$^{46}$& $-$\\ 
QR2	& $-$0.1\p0.4& $-$0.4\p0.2&$-$0.3\p0.3&$-$0.8\p0.7 & $-$ & 0.9$\times$10$^3$& 0.065&1.6$\times$10$^5$& 2.1$\times$10$^{46}$& $-$\\
QR3	& $-$0.2\p0.5& $-$0.3\p0.3&$-$0.3\p0.4&$-$2.0\p1.0 & $-$ & 0.9$\times$10$^3$& 0.030&1.1$\times$10$^5$&1.0$\times$10$^{46}$& $-$\\
QR4	& +1.4\p1.0& $-$& $-$& $-$& 1.5$\times$10$^{-4}$ & 2.7$\times$10$^3$& 0.008&4.8$\times$10$^5$&7.9$\times$10$^{45}$& F270S\\
QR5	&+0.8\p0.6& +1.0\p0.2& +0.9\p0.3 & 0.1\p0.1\tablenotemark{d}& 1.1$\times$10$^{-4}$ & 3.0$\times$10$^3$\tablenotemark{d}& 0.017\tablenotemark{d}&7.1$\times$10$^5$\tablenotemark{d}&1.8$\times$10$^{46}$\tablenotemark{d}& F241\\
QR6	&+0.5\p0.8&+0.4\p0.4&+0.4\p0.4&$-$& 3.7$\times$10$^{-5}$ &$-$&$-$&$-$&$-$ & F257\\
QR7	&+2.0\p0.8&$-$&$-$&$-$& 9.5$\times$10$^{-5}$ & 2.7$\times$10$^3$& 0.005&4.1$\times$10$^5$&4.8$\times$10$^{45}$& F231\\
QR8	& $>$+0.3&+0.7\p0.4&$>$+0.6&$-$& 4.3$\times$10$^{-5}$ &$-$&$-$&$-$&$-$ & F320\\
QR9	& $>$+0.2& $-$&$-$&$-$& 3.2$\times$10$^{-5}$ &$-$&$-$&$-$&$-$ & F362\\
Pistol  &+0.1\p0.8&+0.0\p0.4&+0.1\p0.4&+0.0\p0.6& 2.7$\times$10$^{-4}$ &$-$ &$-$ &$-$ &$-$& F134\\
\enddata
\tablenotetext{a}{S $\propto$ $\nu^{\alpha}$}
\tablenotetext{b}{Derived from the 22.5 GHz flux density where possible; otherwise from the 8.5 GHz flux density (see $\S$4.1.1).}
\tablenotetext{c}{from Figer et al. (1999).}
\tablenotetext{d}{Calculated for extended emission surrounding QR5 using the 22.5 GHz flux density and deconvolved size.}
\end{deluxetable}

\begin{deluxetable}{lcccccccc}
\tabletypesize{\tiny}
\tablecaption{Observed Properties of the Radio Sources in the Arches Cluster}
\tablehead{\colhead{Source} &
\multicolumn{2}{c}{Source Position\tablenotemark{a}} &
\multicolumn{2}{c}{Flux Density (mJy)\tablenotemark{b}} &
\colhead{Spectral}&
\colhead{\mdot}&
\colhead{Near-IR}&
\colhead{Deconvolved}\\
\cline{2-3}
\cline{4-5}
\colhead{Name}&
\colhead{R.A. (J2000)} &
\colhead{Decl. (J2000)} &
\colhead{8.5 GHz} &
\colhead{22.5 GHz} &
\colhead{Index\tablenotemark{c}}&
\colhead{(\msun~\yr)\tablenotemark{d}}&
\colhead{Counterpart\tablenotemark{e}}&
\colhead{Size (\arcsec)}}
\tablecolumns{9}
\startdata
AR1	&17 45 50.42&$-$28 49 22.3	&1.90\p0.05\tablenotemark{\dagger}&1.49\p0.27&$-$\tablenotemark{f}&2.2$\times$10$^{-4}$&F6, N8, C8& 0.14 $\times$0.10\\
AR2	&17 45 50.39&$-$28 49 21.3	&0.25\p0.03&0.60\p0.09&+0.9\p0.2 & 4.6$\times$10$^{-5}$ &F8, N7, C6&ps\\
AR3	&17 45 50.20&$-$28 49 22.3	&0.12\p0.03\tablenotemark{\dagger}&$<$0.25&$>$0.8&1.9$\times$10$^{-5}$&F1, N4, C9&ps\\
AR4	&17 45 50.47&$-$28 49 19.5	&0.34\p0.03\tablenotemark{\dagger}&0.61\p0.09&+0.6\p0.2&4.8$\times$10$^{-5}$&F37, N10, C5&ps\\
AR5	&17 45 50.57&$-$28 49 17.5	&0.16\p0.03&$<$0.25&$>$0.5&1.9$\times$10$^{-5}$ &F4, N11,C2&ps\\
AR6	&17 45 49.76&$-$28 49 26.0	&0.25\p0.03&$<$0.25&$>$0.0&3.0$\times$10$^{-5}$&F19&ps\\
AR7	&17 45 50.83&$-$28 49 26.4	&0.23\p0.03&0.57\p0.09&+0.9\p0.2&4.3$\times$10$^{-5}$&F3, N14, C11&ps \\
AR8	&17 45 50.45&$-$28 49 31.9	&0.16\p0.03\tablenotemark{\dagger}&0.48\p0.09&+1.1\p0.3&3.7$\times$10$^{-5}$&F5, N9&ps\\
AR9	&17 45 50.47&$-$28 49 17.9	&0.15\p0.04\tablenotemark{\dagger\dagger}&$<$0.25&$>$0.5&1.9$\times$10$^{-5}$&F18&ps\\
AR10	&17 45 49.69&$-$28 49 25.9	&0.06\p0.03\tablenotemark{\dagger\dagger}&$<$0.25&$>$1.5&1.9$\times$10$^{-5}$&F2, N1, C13&ps\\
\enddata
\tablenotetext{a}{Positional errors are \p0.01 in RA and \p0.1\arcsec~in DEC.}
\tablenotetext{b}{Flux densities are measured from images which have been corrected for primary beam attenuation.}
\tablenotetext{c}{S $\propto$ $\nu^{\alpha}$}
\tablenotetext{d}{Derived from the 22 GHz flux density where possible; otherwise from the 8.5 GHz flux density (see $\S$4.1.1).}
\tablenotetext{d}{Stellar Identification ``F'' from Figer et al. (2002), ``C'' from Cotera et al. (1996), ``N'' from Nagata et al. (1995).}
\tablenotetext{f}{At 22.5 GHz, AR1 may be missing some extended flux; therefore, a spectral index is not calculated.}
\tablenotetext{\dagger}{Flux density has changed by 12-30\% from Lang et al. (2001b)}
\tablenotetext{\dagger\dagger}{New sources detected in these observations.}
\end{deluxetable}

\begin{deluxetable}{lccccc}
\tabletypesize{\small}
\tablecaption{Properties of Clusters of Radio Sources in Regions of Recent Star Formation}
\tablehead{
\colhead{}& 
\colhead{Bolometric} &
\colhead{Number of} & 
\colhead{Most Luminous} &
\colhead{Diameter} &
\colhead{} \\ 
\colhead{Region} &
\colhead{Luminosity ($L_\odot$)} &
\colhead{Members}  &
\colhead{Member (mJy~$\cdot~(\frac{d}{kpc})^2$)\tablenotemark{a}}  &
\colhead{(pc)}&
\colhead{Reference}}
\tablecolumns{6} 
\startdata
Orion & $\sim$2$\times$10$^5$  & 77 & 8.8 & 0.3 & Zapata et al. (2004)  \\
NGC~1579 & $\sim$2$\times$10$^3$ & 16 & 0.6 & 0.3 & Stine \& O'Neal (1998) \\
GGD~14 & $\sim$10$^4$ & 6 & 0.2 & 0.2 & G\'omez et al. (2002)  \\
NGC~1333 & $\sim$120 & 44 & 0.3 & 0.7 & Rodr\'\i guez et al. (1999) \\
NGC 2024 & $\sim$5$\times$10$^4$  & 25 & 3.0 & 0.5 & Rodr\'\i guez et al. (2003) \\
W40      & $\sim$5$\times$10$^4$  & 14 & 2.8 & 0.2 & Rodr\'\i guez \& Reipurth (2005) \\
Quintuplet & $\sim$ 10$^8$ & 10 & 218 & 2.0 & This paper \\
Arches & $\sim$10$^8$ & 10 & 122 & 0.7 & This paper \\
NGC3603 & $\sim$5$\times$ 10$^7$ & 2 & 147 & 0.7 & Moffat et al. (2002)\\
NGC3603\tablenotemark{b} &$\sim$ 5$\times$ 10$^7$ & 4 & 687& 5.0 & M\"{u}cke et al. (2002)\\
\enddata
\tablenotetext{a}{8.5 GHz flux density times distance in kpc squared}
\tablenotetext{b}{Sources out to a 5.0 pc radius from cluster core.}
\end{deluxetable}

\begin{figure}
\plotone{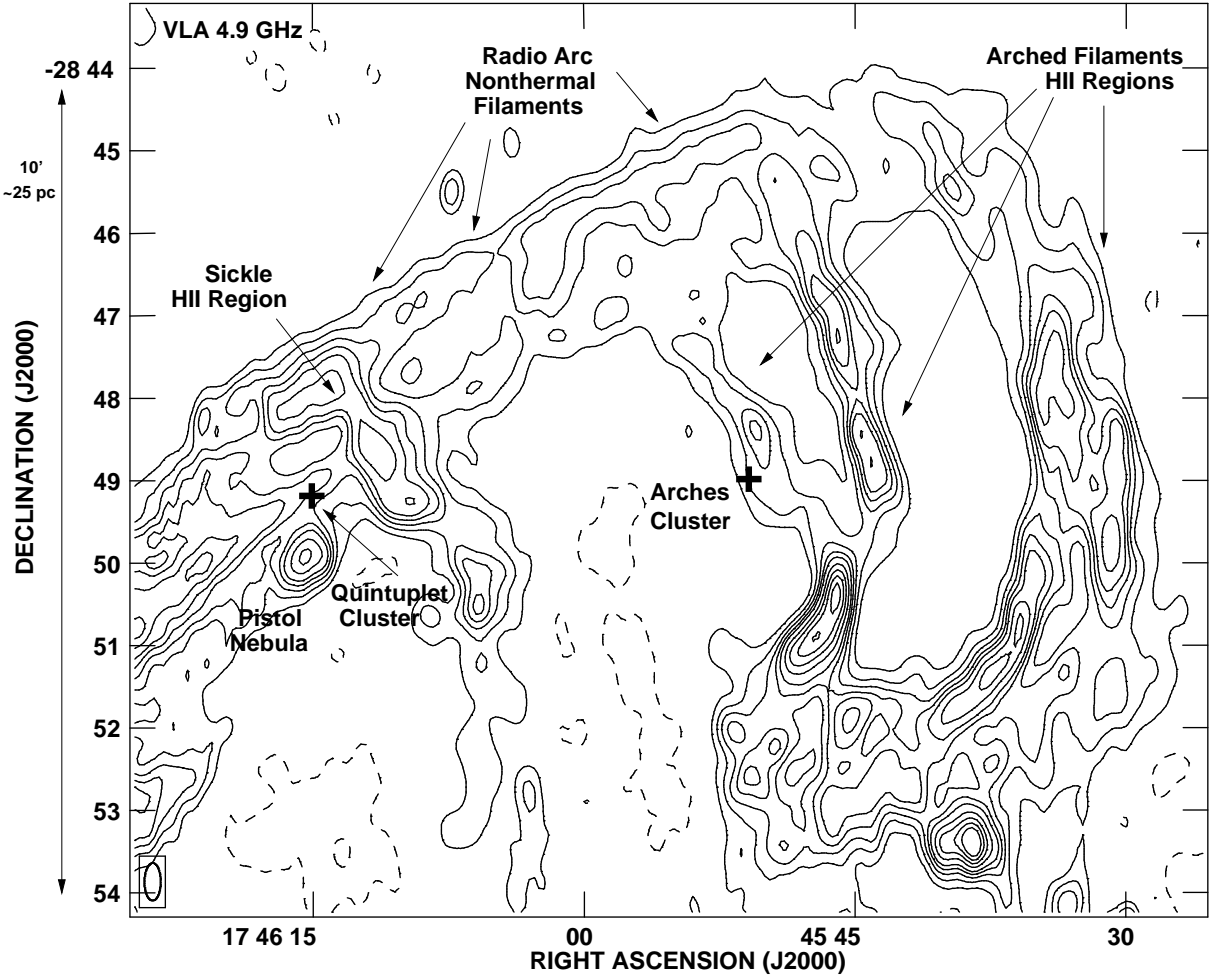}
\caption{VLA 4.9 GHz continuum image of the Galactic center Radio Arc region (located $\sim$30 pc in projection from SgrA$^*$), highlighting the interstellar environment and location of the Arches and Quintuplet clusters. The Sickle \ion H2 region, Pistol nebula, and Arched Filament \ion H2 regions are prominent features in this region. The linear nonthermal filaments of the Radio Arc are also apparent in this image. The resolution of this image is 27.4\arcsec $\times$ 11.9\arcsec, PA=-0.3\arcdeg~(the D-array configuration of the VLA) and is from Lang {\it (in preparation)}. The image is a mosaic of five fields and has been corrected for primary beam attenuation. Contours represent levels of -30, 30, 60, 90, 120, 150, 200, 250, 300 and 350 mJy beam$^{-1}$.\label{fig1}}
\end{figure}

\begin{figure}
\plotone{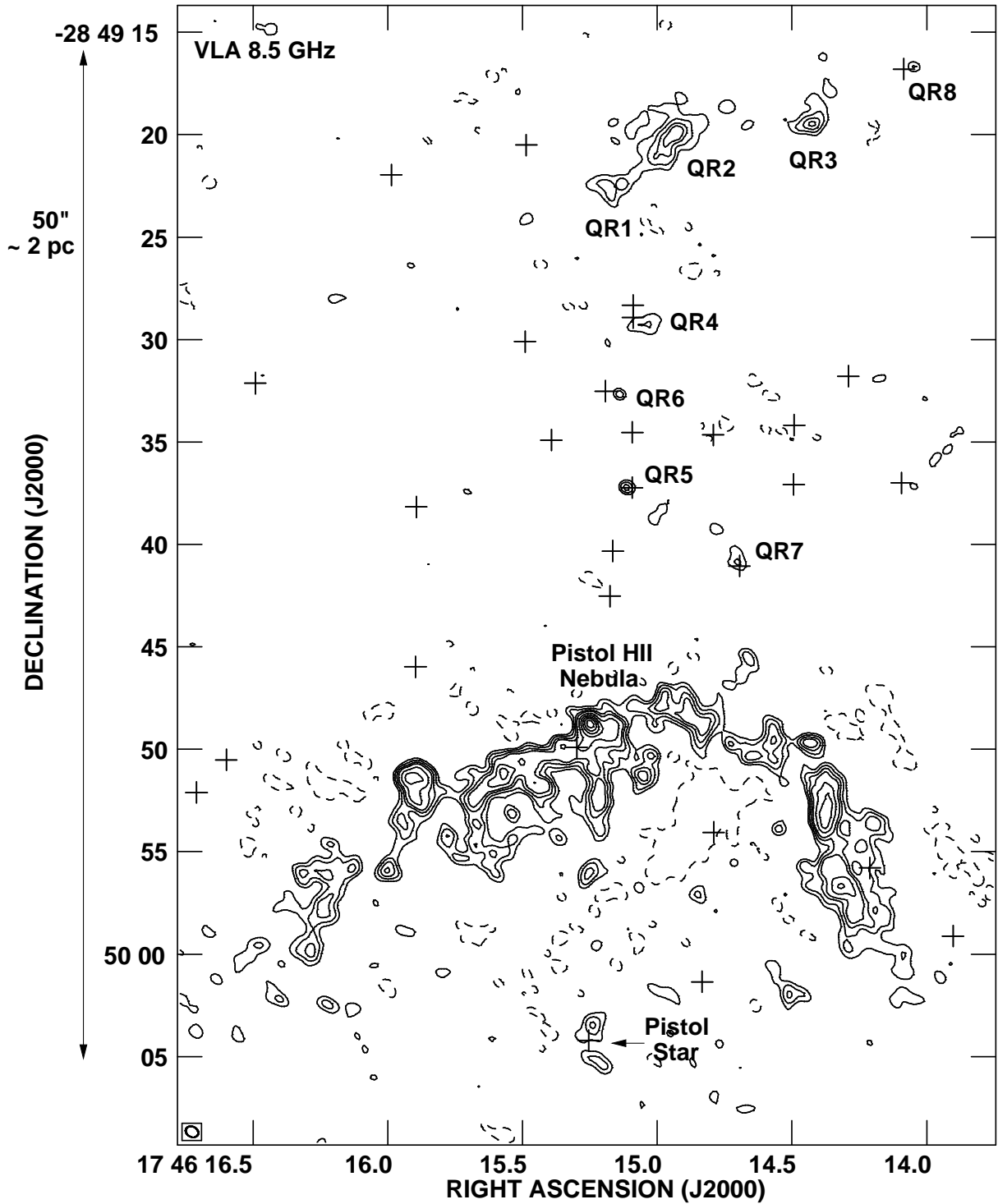} 
\caption{VLA 8.5 GHz continuum image of the Quintuplet cluster. The resolution is 0.64\arcsec$\times$0.50\arcsec, PA=62\arcdeg, and the rms noise level is 55 $\mu$Jy beam$^{-1}$. The image has not been corrected for primary beam attenuation. Contours represent 3, 5, 7.5, 10, 15, 20, 25, 30, 35, 40, 45, and 50 times the rms noise level. The radio sources QR1-8 are labelled in the figure. QR9 is located $\sim$ 30\arcsec~to the NE of the center of this image. The crosses represent the positions of stars in the Quintuplet cluster, which have been classified by Figer et al. (1999) as evolved, massive stars. A systematic shift of $\Delta$RA=0.9\arcsec~was introduced to align the radio and {\it HST} near-IR images. 
\label{fig2}}
\end{figure}

\begin{figure}
\begin{center}
\includegraphics[scale=0.75]{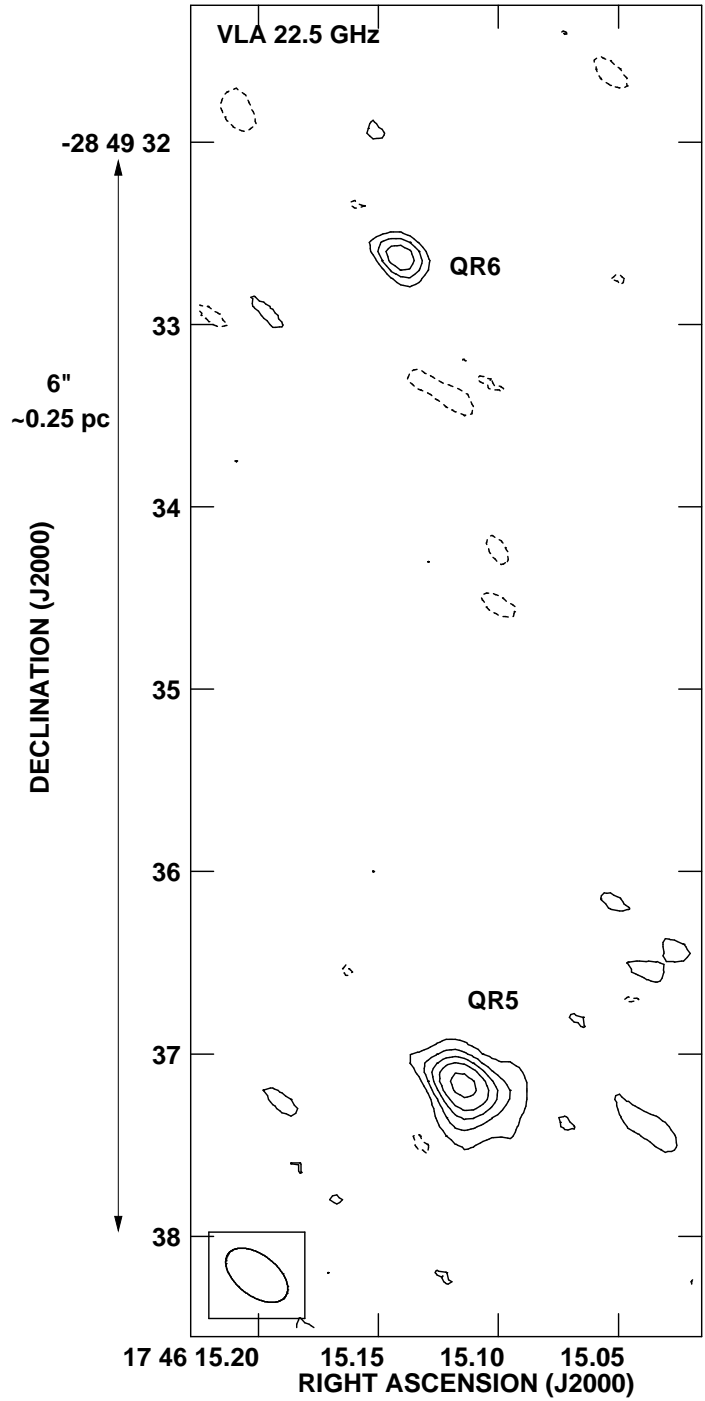}
\caption{VLA 22.5 GHz image of the compact sources QR5 and QR6 from the ``high'' resolution 22.5 GHz image of the Quintuplet Cluster. The resolution of this image is 0.39\arcsec$\times$0.22\arcsec, PA=52.7\arcdeg. The rms noise level in this image is 65 $\mu$Jy beam$^{-1}$ and the contours correspond to -3, 3, 5, 7, 9 and 12 times this level. This image has not been corrected for primary beam attenuation. 
\label{fig3}}
\end{center}
\end{figure}

\begin{figure}
\begin{center}
\includegraphics[scale=0.60]{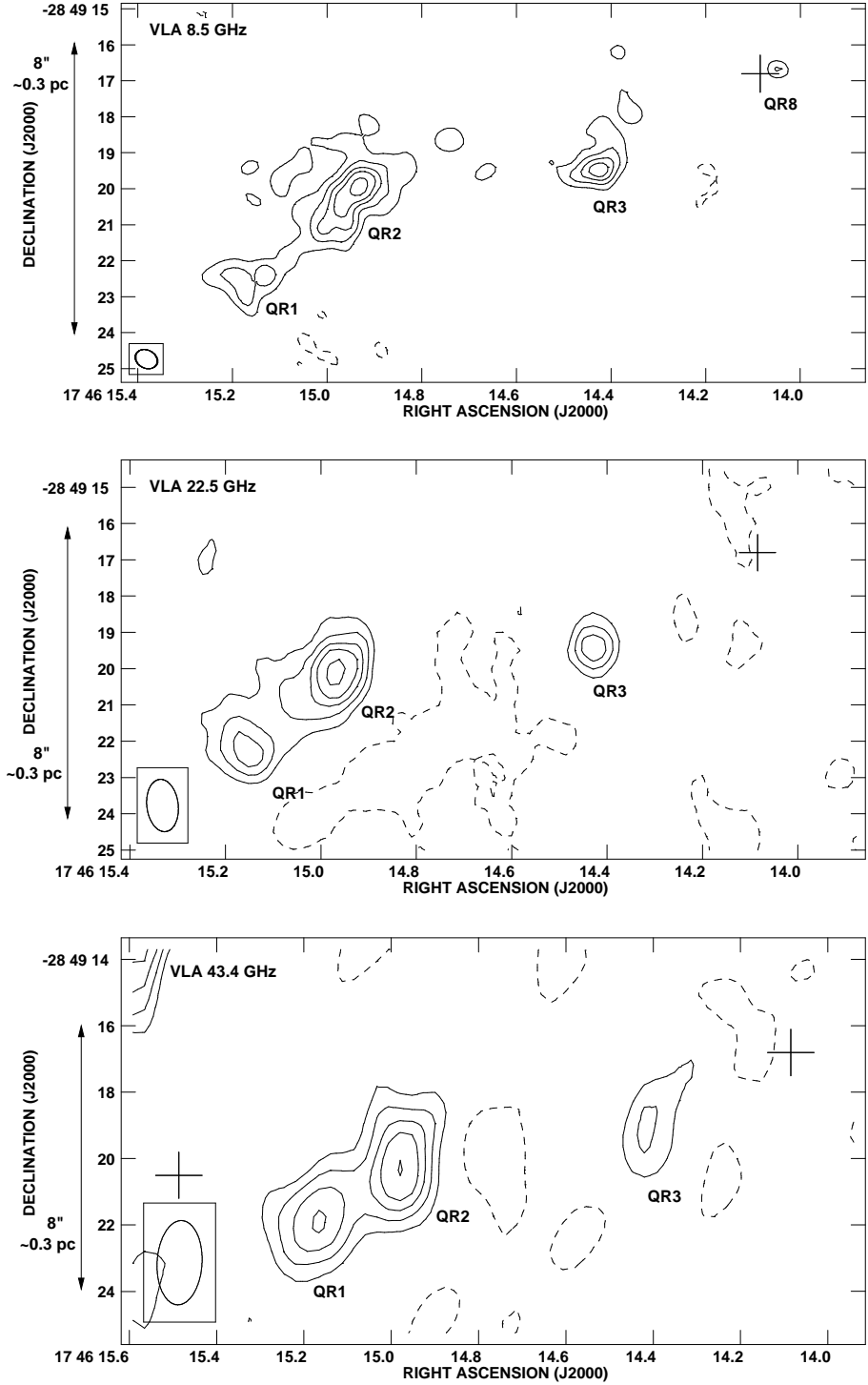}
\caption{VLA images at 8.5 GHz (top), 22.5 GHz (middle) and 43.4 GHz (bottom) of the extended radio sources to the North of the Quintuplet cluster, QR1, QR2 and QR3. The compact source QR8 is clearly detected in the top panel and its stellar counterpart is indicated by a cross (as in Figure 2, a systematic shift of $\Delta$RA=0.9\arcsec~was introduced to align the radio and {\it HST} near-IR images). The 8.5 GHz image (top) is an inset from Figure 2, with a resolution of 0.64\arcsec$\times$0.50\arcsec, PA=62\arcdeg; the 22.5 GHz image (middle) is the ``low'' resolution image and has a resolution of 1.45\arcsec$\times$0.87\arcsec, PA=6.8\arcdeg; the 43.4 GHz (bottom) image has a resolution of 2.54\arcsec$\times$1.35\arcsec, PA=3.3\arcdeg. Contours in each case correspond to -3, 3, 5, 7, 9, and 12 times the rms levels of 55 $\mu$Jy beam$^{-1}$ (8.5 GHz), 75 $\mu$Jy beam$^{-1}$ (22.5 GHz) and 100 $\mu$Jy beam$^{-1}$ (43.4 GHz). None of these images have been corrected for primary beam attenuation.    
\label{fig4}}
\end{center}
\end{figure}

\begin{figure}
\plotone{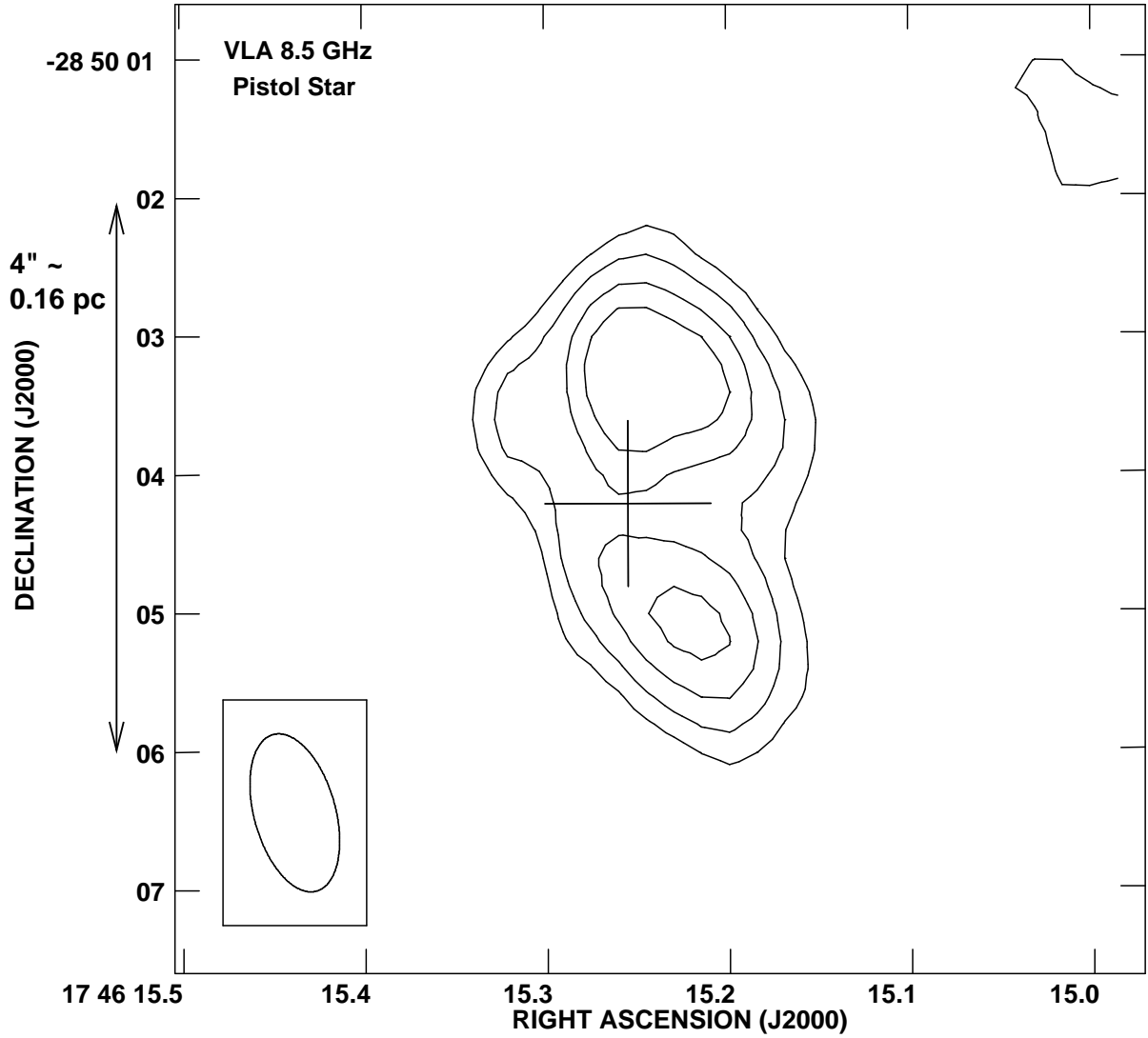}
\caption{VLA 8.5 GHz image of the Pistol star nebula from an additional observation on the Pistol star (see Tables 1 and 2). The resolution here is 1.17\arcsec$\times$0.59\arcsec, PA=15.2\arcdeg~and contour levels represent 0.18, 0.3, 0.45, and 0.6 mJy (which correspond to 1.8, 3, 4.5, and 6 times the rms level). The cross represents the position of the Pistol star from Figer et al. (1999) and its size represents the 1\arcsec~positional alignment error between the {\it HST/NICMOS} and VLA reference frames.  This image has not been corrected for primary beam attenuation. 
\label{fig5}}
\end{figure}

\begin{figure}
\plotone{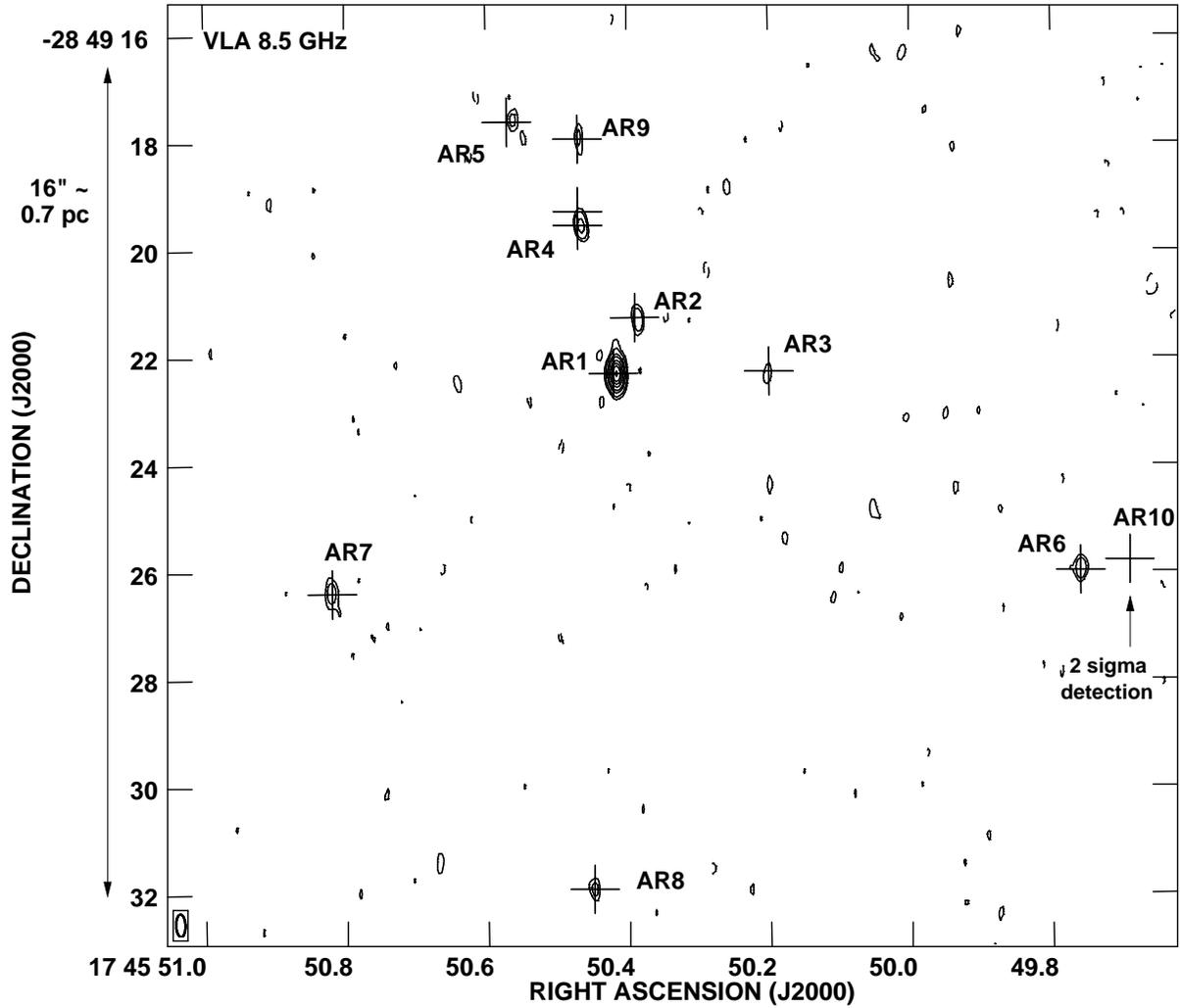}
\caption{VLA 8.5 GHz image of the Arches cluster showing the ten compact sources, AR1-10. The resolution of the image is 0.42\arcsec$\times$0.17\arcsec, PA=1.6\arcdeg, with an rms noise level of 25 $\mu$Jy beam$^{-1}$. The contour levels represent -0.075, 0.075, 0.125, 0.250, 0.375, 0.625, 0.875, 1.125, 1.375, and 1.5 mJy beam$^{-1}$ (which correspond to -3, 3, 5, 10, 15, 25, 35, 45, 55, and 60 times the rms level). The crosses represent the positions of stars in the Arches cluster, which have been classified by Figer et al. (2002) as evolved, massive stars based on their near-IR emission lines. A systematic shift of $\Delta$RA=0.8\arcsec, $\Delta$DEC=0.5\arcsec was introduced to align the radio and near-IR {\it HST} images. The sizes of the crosses ($\sim$1\arcsec) represents this positional error in alignment. This image has not been corrected for primary beam attenuation.  
\label{fig6}}
\end{figure}

\end{document}